\RequirePackage{fix-cm}
\documentclass{svjour3}                     
\smartqed  
\usepackage{graphicx}
\usepackage{hyperref}
\usepackage[export]{adjustbox}
\usepackage{enumitem}
\usepackage{makecell}
\setlist{nosep}

\newcommand{\bi}{\begin{itemize}[leftmargin=*]}
\newcommand{\ei}{\end{itemize}}
\newcommand{\be}{\begin{enumerate}[leftmargin=*]}
\newcommand{\ee}{\end{enumerate}}

\usepackage[skins]{tcolorbox}
\usepackage{subcaption}
\usepackage{multirow}
\newenvironment{RQ}{\vspace{2mm}\begin{tcolorbox}[enhanced,width=\linewidth,size=fbox,colback=blue!5,drop shadow southeast,sharp corners]}{\end{tcolorbox}}
\begin{document}

\title{Approaching Code Search for Python as a Translation Retrieval Problem with Dual Encoders}


\titlerunning{Code Search with Dual Encoders}        

\author{Monoshiz Mahbub Khan \and
        Zhe Yu
}


\institute{M. Khan \at
              Rochester Institute of Technology\\
              Rochester, New York\\
              \email{mk7989@rit.edu}\\
           Z. Yu \at
              Rochester Institute of Technology\\
              Rochester, New York\\
              \email{zxyvse@rit.edu}
}


\maketitle

\begin{abstract}
Code search is vital in the maintenance and extension of software systems. Past works have used separate language models for the natural language and programming language artifacts on models with multiple encoders and different loss functions. Similarly, this work approaches code search for Python as a translation retrieval problem while the natural language queries and the programming language are treated as two types of languages. By using dual encoders, these two types of language sequences are projected onto a shared embedding space, in which the distance reflects the similarity between a given pair of query and code. However, in contrast to previous work, this approach uses a unified language model, and a dual encoder structure with a cosine similarity loss function. A unified language model helps the model take advantage of the considerable overlap of words between the artifacts, making the learning much easier. On the other hand, the dual encoders trained with cosine similarity loss helps the model learn the underlining patterns of which terms are important for predicting linked pairs of artifacts. Evaluation shows the proposed model achieves performance better than state-of-the-art code search models. In addition, this model is much less expensive in terms of time and complexity, offering a cheaper, faster, and better alternative.

\end{abstract}

\section{Introduction}
\label{Introduction_section}
As defined by Husain et al. \cite{husain2019codesearchnet}, ``Semantic code search is the task of retrieving relevant code given a natural language query". In other words, the goal of code search is, given a natural language artifact or sequence, find the corresponding programming language artifact or sequence from a pool of available or possible artifacts. In practical scenarios, there might be multiple matching programming language artifacts for a natural language artifact counterpart. The goal there might be vary from picking out the most relevant one, or simply retrieving them all. Retrieving these target artifacts for a given query can help implement specific features more conveniently, find software libraries for specific functionality, navigate codebases, or even retrieve pieces of code that need some form of modification \cite{salza2022effectiveness}. However, manually retrieving these target artifacts for existing systems and artifacts can be tedious and expensive \cite{rath2018traceability,lin2021traceability}. Moreover, traditional information retrieval processes do not perform well in code search, due to the different languages used in the queries and code--- ``there is often little shared vocabulary between search terms and results" \cite{husain2019codesearchnet}. In addition, real life data concerning programming language artifacts may sometimes include variable or function names that may not be as easy to understand the semantic of only from the names \cite{guo2021graphcodebert}. As a result, an overlap of vocabulary between artifacts does not necessarily guarantee better training or learning, especially for large datasets where such an overlap occurring between unrelated artifacts might be commonplace.

Contemporary works in code search have approached this as a retrieval task, where there is already a pool of possible target artifacts, and the model simply has to learn which one to pick for a given query. More recent works have utilized separate encoders for each type of artifact, similar to translation tasks. This, in turn, has framed the code search task as a translation retrieval task.
The CodeSearchNet challenge \cite{husain2019codesearchnet} is perhaps the most significant past work relevant to the task of code search. This work introduces the CodeSearchNet dataset, consisting of documentation-function pairs from six programming languages, and trains several models to retrieve the programming language artifact most likely to be linked to a given natural language pair. Feng et al. \cite{feng2020codebert} build upon this task by introducing more complex structures with some levels of pre-training, using the same dataset for the task. Lu et al. \cite{lu2021codexglue} expand on this by using a subset of the CodeSearchNet dataset on the same models. Guo et al. \cite{guo2021graphcodebert} incorporate data flow information from the programming language artifacts into its training process, while Wang et al. \cite{wang2021syncobert} utilize Abstract Syntax Tree (AST) representations of the programming language artifacts. Wang et al. \cite{wang2023codet5+} pre-train a single model on separate tasks with both different unimodal and bimodal data as different steps to ensure the model can be flexibly used in different modes for different tasks.

\begin{table}[t]
    \centering
    \caption{Example pairs with overlapping words or partial words}
    \begin{tabular}{p{0.1\linewidth}|p{0.3\linewidth}|p{0.5\linewidth}}
    \hline
         Linked & Text & Code\\ \hline
         Yes & \underline{Decompresses} data for Content-Encoding: \underline{deflate}.\newline (the \underline{zlib} \underline{compression} is used.) & def un\underline{deflate}(data):\newline
         import \underline{zlib} \newline
         \underline{decompress}obj = \underline{zlib}.\underline{decompress}obj(-\underline{zlib}.MAX\_WBITS) \newline
         return \underline{decompress}obj.\underline{decompress}(data) +\newline
         \underline{decompress}obj.flush()\newline\\ \hline

         Yes & Checks if a \underline{task} is either \underline{queued} or running in this executor\newline :param \underline{task\_instance}: \underline{TaskInstance}\newline
        :return: True if the \underline{task} is known to this executor\newline & 
        def has\_\underline{task}(self, \underline{task\_instance}):\newline
        if \underline{task\_instance}.key in self.\underline{queued}\_\underline{task}s or \underline{task\_instance}.key in\newline
        self.\underline{running}:\newline
            return True\newline\\ \hline
         No  &  Dumps a \underline{database} \underline{table} into a tab-delimited \underline{file}  &
         def bulk\_load(self, \underline{table}, tmp\_\underline{file}):\newline
        self.copy\_expert("COPY {\underline{table}} FROM STDIN".format(\underline{table}=\underline{table}), tmp\_\underline{file})  \\ \hline
    \end{tabular}
    \label{overlap_example}
\end{table}

\subsection{Proposed approach}
\label{proposed_approach}
The work here considers the same scenario as the CodeSearchNet challenge for Python, where there are pools of natural language artifacts and programming language artifacts, and every natural language query has at least one correct or linked target programming language artifact. This work builds upon the upon mentioned past works for this retrieval task. This work incorporates a unified language model to generate word embeddings, and utilizes a dual encoder model alongside a cosine similarity based loss function to learn from these embeddings.

One of the primary components of the work here is the use of a unified language model to generate the word embeddings for both the natural and programming language artifacts. In Table-\ref{overlap_example}, the first two examples show two linked artifact pairs. These linked pairs demonstrate a significant overlap in words or between them. This pattern of overlapping words or parts of words is consistent for linked pairs throughout the data. In other words, linked artifacts in the data show a shared vocabulary between them. A unified language model is able to take advantage of a pair's shared vocabulary to generate similar embeddings for the artifacts. Therefore, the nature of the data directly motivates the use a unified language model for the word embeddings. The FastText word embeddings used here can leverage sub-word level information, making similar artifacts' embeddings similar.

However, there are also non-linked artifact pairs presenting in the data which also display a significant overlap of words, as it can be seen in the negative example in Table-\ref{overlap_example}. Therefore, it is necessary to learn which artifacts are actually linked and which are not, regardless of their embeddings' initial similarity. The dual encoders trained with cosine similarity-based loss are used to achieve that goal. The initial embeddings of the natural language and programming language artifacts are projected onto a shared embedding space by the trained language-specific encoders. In this shared embedding space, linked pairs are closer to each other than non-linked pairs when measured with cosine distance. In this way, the similarity between any pair of natural language query and programming language artifact can be predicted by their cosine similarity score in the shared embedding space. This lends the learned model the capability of learning which terms are more important and distinguishing between linked pairs and non-linked pairs. In contrast to the use of classification based loss functions in previous work, the cosine similarity loss is more appropriate for this task and consequently performs better empirically.

\subsection{Research questions} This work evaluates the proposed model on the CodeSearchNet dataset and some of its variations, and focusing only on Python-based code. Following different pre-processing steps, this data is fed to our model for the code search task. Detailed steps and processes are described in a later section. The following research questions will be explored in the rest of this paper:

\bi
    \item \textbf{RQ1:  Is a simpler dual encoder architecture with unified word embeddings and cosine 
                        similarity based loss function more effective at code search in Python?}
                        Dual encoder architectures have shown to be effective on the code search task in the past. Using two encoders helps distribute the training between each encoder, and makes their learning more efficient and thorough. Past work has incorporated different contextual information into this training to improve performance. Most of these works use separate language models, alongside pre-trained models for their word embeddings, but also use contextual information gathered from the data separately. In this RQ, we will explore whether the proposed approach with (i) a single language model trained on all of the data to generate the word embeddings, and (ii) a cosine similarity based loss function used to teach the model of each pairs' similarities and differences rather than using a classification loss function to classify between both linked and non-linked pair, outperforms existing approaches in searching for Python code.
                    
    \item \textbf{RQ2:  What roles do the data and the models play in making the code search task more 
                        effective?}
                        The datasets used in this work are comprised of text and code pairs from real projects across different tasks, all based on Python. The other key goal of this work is to investigate how the nature of this varied and diverse data affects the model and the task, and vice versa.
\ei

\subsection{Contributions}
\label{Contributions}

The contributions of this work are:
\begin{itemize}
    \item This work builds a pipeline centered around the dual encoder architecture with trained unified FastText word embeddings and cosine similarity based loss function that considers the semantic similarity between each pair of artifacts for the code search task.  
    \item Evaluation shows that the proposed approach outperformed state-of-the-art approaches on the CodeSearchNet Python dataset, the AdvTest dataset, and the DGMS dataset.
    \item Analysis on the results show how important an overlap of words between paired artifacts is for the model to correctly learn of pairs' similarities and dissimilarities.
    \item Analysis also proves the model being capable of learning which pairs are relevant, even if function or variable names in common between artifact pairs are transformed.
    \item The code and data used in this work are made publicly available at \url{https://github.com/hil-se/CodeSearch}.
\end{itemize}

The following sections go into more detail on the background and related work, the data and methods used in this work, the results and their implications, and possible future work.

\section{Background and Related Work}
\label{background}
\subsection{Code Search}
\label{general_code_search}
Looking at earlier contributions to the code search task, Fernandes et al. \cite{fernandes2018structured} treated this task as a summarization problem, and used a sequence encoder in combination with a graph neural network to generate natural language documentation when provided a programming language query. In other words, this work did not retrieve the relevant artifacts for given queries, but rather generated them. While this allowed the flexibility of documentation generation for cases where the data might be incomplete or some artifacts might not have corresponding pairs, this approach might not be ideal where the data is complete with all programming language artifacts having existing natural language counterparts, since the existing documentation might be more appropriate to human judges or users over the ones learned and generated by the model.

Other approaches apply different vectorization for the programming language artifacts and natural language queries to predict with vector space models~\cite{van2017combining}. For example, the NCS model \cite{sachdev2018retrieval} uses the word embeddings to vectorize in an unsupervised model. Cambronero et al. \cite{cambronero2019deep} later proposed their own model, UNIF--- which served as a supervised version and extension of the NCS model. In contrast, Yao et al. \cite{yao2019coacor} uses a framework based on reinforcement learning to first train a model to generate natural language documentation for given programming language artifacts, and includes these documentations with a code retrieval model to retrieve programming language artifacts relevant to a given query.

Looking at past works that made use of other approaches, Zamani et al. \cite{zamani2014noun} used a Noun-Based Feature Location and a time-aware weighting technique to find the location of some source code, given some request for change. In contrast, P{\'e}rez et al. \cite{perez2018automatic} uses Feature Location techniques to reformulate the queries to better find corresponding artifacts. On the other hand, Kevic et al. \cite{kevic2014automatic} reforumlates the natural language queries using certain heuristics, such as the term frequencies, part of speech and more to better find relevant artifacts.

Satter et al. \cite{satter2016search} uses the the terms present in the user's history in conjunction with their query to find the most similar artifacts from its pool. Wang et al. \cite{wang2014active} integrates user feedback to reorder its predictions, refining future predictions. Similarly, Gay et al. \cite{gay2009use} integrate developers' feedback to refine its information retrieval based concept location methods. In contrast, Yu et al. \cite{yu2016apibook} uses the semantic and type information of queries and programming language artifacts to assign similarity scores to each of the retrieved artifacts, and then ranks them. On the other hand, Lemos et al. \cite{lemos2015can} integrates a natural language thesaurus in keyword-based code search interface-driven code search, and compares their performances. Yang et al. \cite{yang2014swordnet} expands on this by using semantically similar word pairs, mining through the training text data and considering the contexts of pairs.

Balachandran \cite{balachandran2015query} converts the queries into Abstract Syntax Trees (AST), and compares the AST and its subtrees against the ASTs and subtrees of the programming language artifacts. Wang et al. \cite{wang2016autoquery} generates Program Dependence Graphs (PDGs), and uses graph mining methods to extract common structures in these PDGs and coverts them into dependency queries to retrieve relevant code snippets.

\subsection{Code Search with LLMs}
\label{codesearch_with_llms}
Large language models (LLMs) have gained traction in recent times due to their ability to produce high quality texts across a variety of natural language processing tasks. And with recent advances in generative AI models such as GPT-4 \cite{achiam2023gpt}, Bing Chat \cite{mehdi2023reinventing} and Claude2 \cite{claude2}, the quality of the generated code from a natural language query has become better and better. However, these generative AI models are still not suitable for the code search task largely due to their black-box nature. As an example, Chai et al. \cite{chai2022cross} found that the pre-trained GPT-2 model \cite{radford2019language} did not perform better than a transferred model with few-shot meta-learning in domain-specific code search.

In addition, a few reasons have prevented the use of LLMs removing the need for code search. Firstly, code search engines are often used internally by organizations maintaining or re-using
their codebase. For maintenance or similar tasks, the retrieved code snippets or functions must be the exact snippets or functions in the codebase. LLMs used to generate code can often generate code that is not an exact match of the existing code, which defeats the purpose of maintenance or updates. Secondly, some of this codebase often contains proprietary code. Code generated by LLMs often cannot be claimed as proprietary, especially with off-the-shelf LLMs, which can often show better performance due to its training on large-scale publicly available data. Finally, code generated by LLMs are often not perfect. Requiring an extra step of verifying that the generated code is correct and usable creates an extra step in the process, requiring more time and effort.

\subsection{Dual Encoders}
\label{dualencoder_background}
\begin{figure*}[t!]
    \centering
    \begin{subfigure}[h!]{0.4\linewidth}
        \centering
        \includegraphics[width=\linewidth, valign=c]{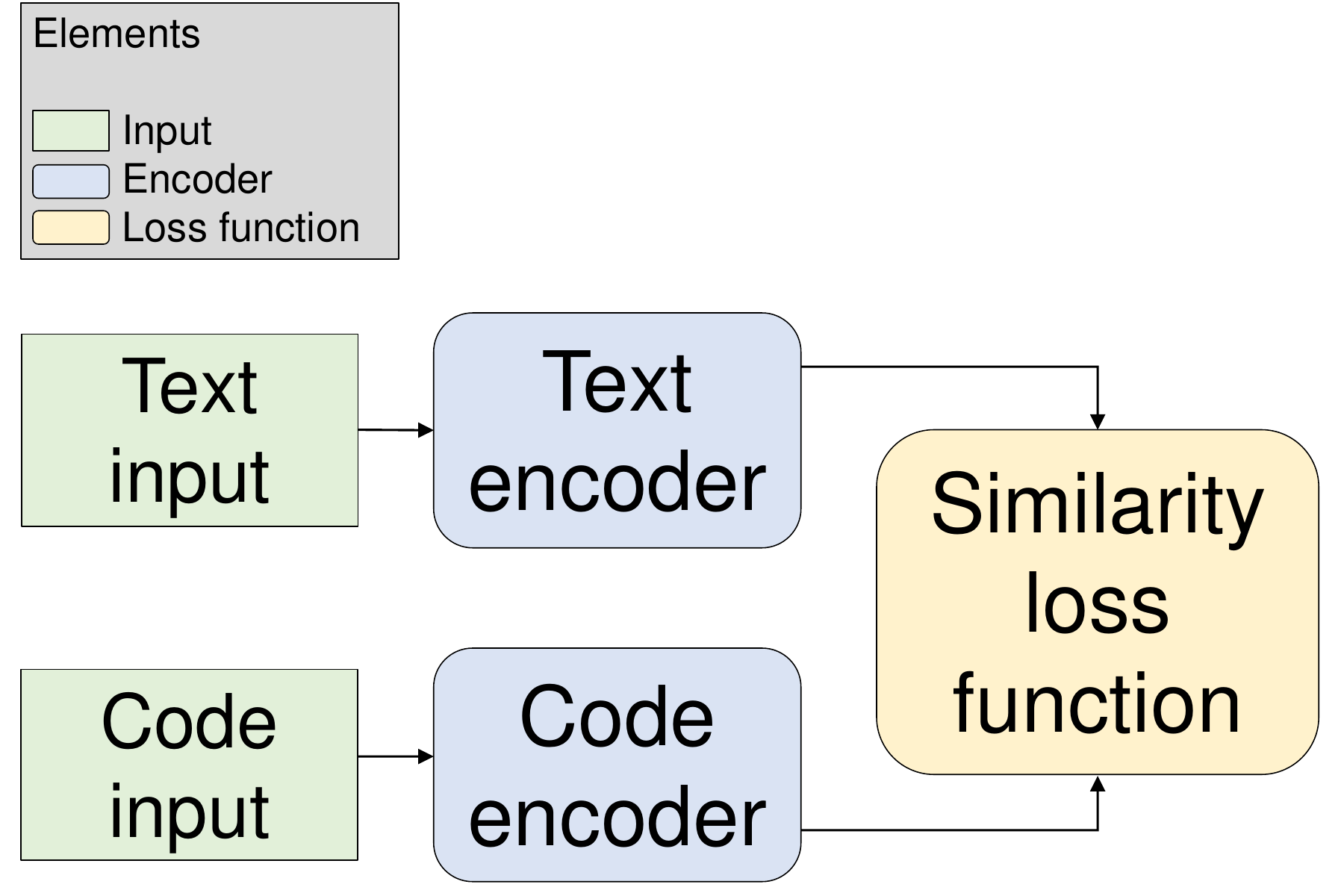}
        \caption{Simplified original Dual-encoder architecture} 
        \label{fig:original_dual_encoder}
    \end{subfigure}%
    ~ 
    \begin{subfigure}[h!]{0.5\linewidth}
        \centering
        \includegraphics[width=\linewidth, valign=c]{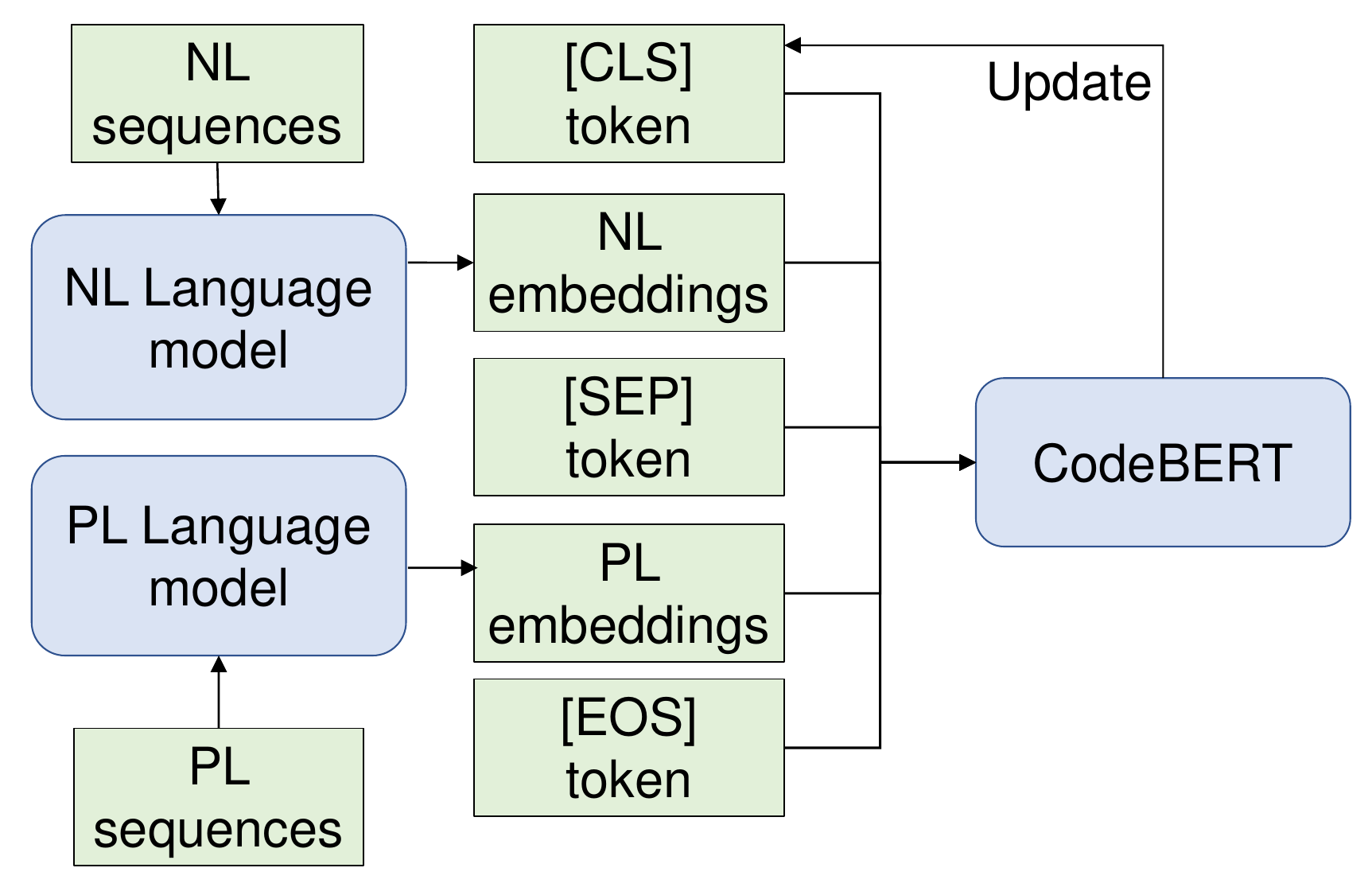}
        \caption{CodeBERT pre-training stage}
        \label{codebert_pretraining}
    \end{subfigure}%
    \hfill
    \begin{subfigure}[h!]{0.5\linewidth}
        \centering
        \includegraphics[width=\linewidth, valign=c]{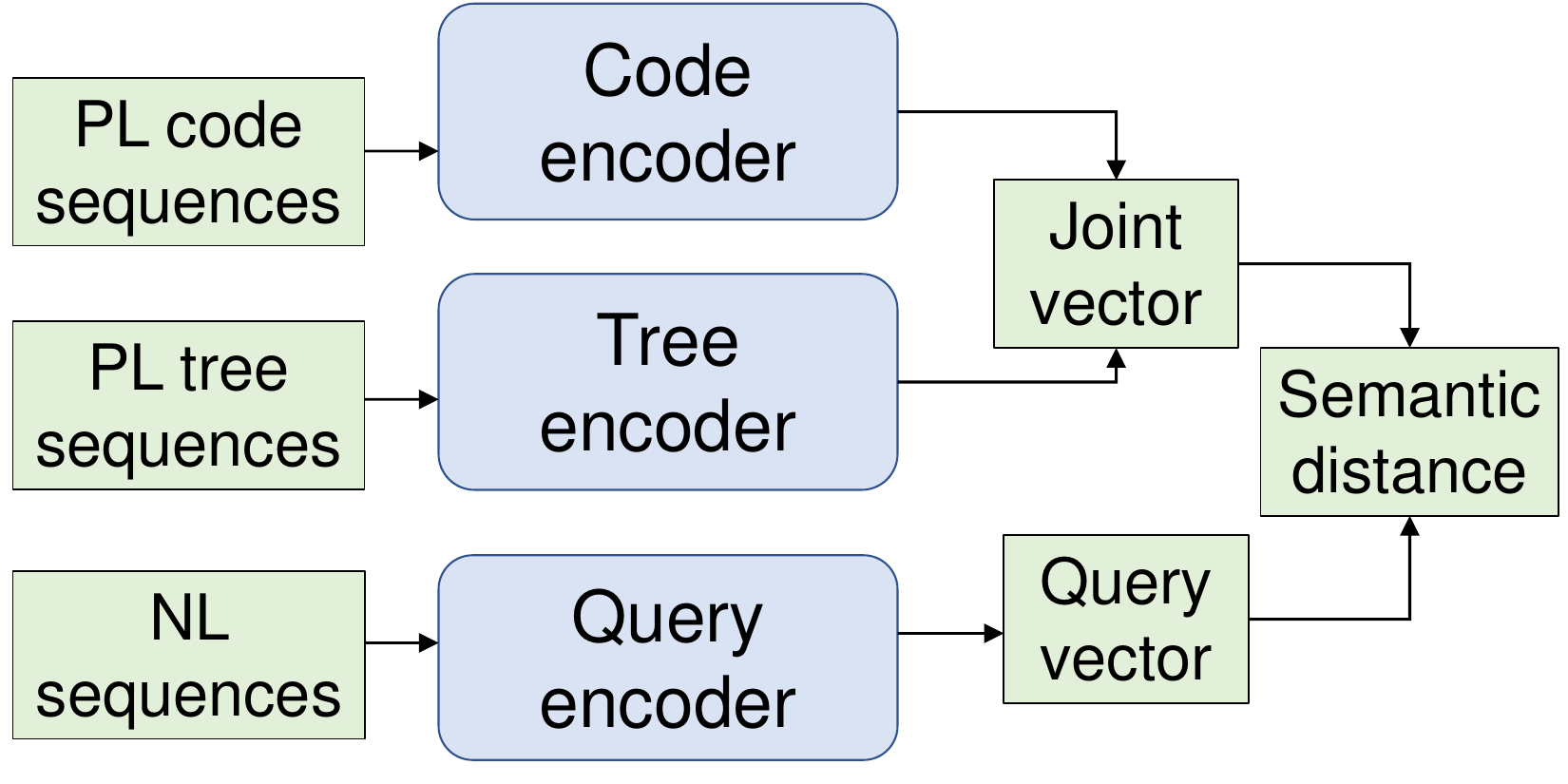}
        \caption{UNI-LCRS training stage}
        \label{unilcrs}
    \end{subfigure}
    \caption{Dual encoder architectures in previous work}
\end{figure*}

Dual-encoder architecture is a machine learning technique originally proposed by Guo et al. \cite{guo2018effective} for bilingual sentence embedding. The goal of a dual-encoder is to encode two different types of inputs into a shared embedding space so that the distance between two embeddings reflects the similarity between two inputs.  In Guo et al.~\cite{guo2018effective}, it was applied to solve the bilingual translation retrieval problem. It trains two separate encoders simultaneously, as shown in Figure-\ref{fig:original_dual_encoder}, to encode sentences from two different languages ($x$ and $y$). The encoded sentences ($u$ and $v$) can then be scored by their dot production to find the most similar sentence pairs. Similar to how the translation task aims to make two sequences' encodings from two languages more similar, the dual encoder structure can be used for code search with an objective of making the encodings of the natural text and programming text sequences similar. The translation retrieval problem can then be modeled as a ranking problem to place $y_i$, the true translation of $x_i$, over all other sentences in $Y$. $P(y_i|x_i)$ can be expressed as a following log-linear model and softmax loss can be used to train the weights of the encoders \cite{yang2019improving}:
\begin{equation}
\centering
P(y_i|x_i) = \frac{e^{S(x_i,y_i)}}{\sum_{y\in Y}{e^{S(x_i,y)}}}
\label{eq:trans}
\end{equation}

\subsection{Code Search with Dual Encoders}
\label{codesearch_dualencoders_subsection}
CODEnn \cite{gu2018deep} was one of the earlier works to utilize multiple encoders for code search. This model uses separate RNN networks for the method names from the code snippet and an API sequence from the code snippet, as well as using a feed forward network on the code tokens' sequence. The model also uses an RNN network to encode the natural language query sequence. The encoded and summarized programming language sequence and the encoded natural language query sequences are used to calculate cosine similarity scores, calculate the loss, and update the weights. In contrast, SCS \cite{husain2018create} uses a sequence-to-sequence GRU network to generate natural language sequences given a programming language sequence, and trains an LSTM network as a language model on only the natural language sequences. Afterwards, the programming language encoder portion of the sequence-to-sequence network and the language model are used together with added components on programming language and natural language sequences. These encoded sequences are then used to calculate a cosine similarity score for the sequences.

Husain et al. \cite{husain2019codesearchnet} introduced the CodeSearchNet dataset, a large and diverse collection of data containing pairs of natural text in the form of comments, documentation and description, and programming language text, in the form of fully coded functions from six different programming languages. Several subsequent works have trained and evaluated their models on this dataset. Husain et al. \cite{husain2019codesearchnet} used two encoders on this dataset for the trace link retrieval task.

Lin et al. \cite{lin2021traceability} worked on trace link prediction on the OSS dataset, but also evaluated their proposed model, T-BERT on the Python division of the CodeSearchNet corpus \cite{husain2019codesearchnet} for the code search task. An aspect that helped these models perform better than previous work was the process of generating pairs of non-linked artifacts as negative examples for their models to learn from. Another key factor that helped achieve high performance for this task, was using word embeddings generated by the CodeBERT model \cite{feng2020codebert}, rather than building or training a new model for embeddings from scratch.

Gu et al. \cite{gu2021multimodal} utilizes AST (Abstract Syntax Tree) representations of the programming language sequences on the code search task. This work trains separate encoders on the programming language sequences, the natural language sequences, and on the AST representation sequences, and combines the programming language-related sequences. The model then uses this joint sequence encodings and the natural language encodings to calculate the cosine similarity. In contrast, Ling et al. \cite{ling2021deep} represents the natural language and programming language sequences in graph formats before using separate RGCNs (Relational Graph Convolutional Network) to encode these sequences and calculate their cosine similarity scores. The CoCoSoDa model \cite{shi2023cocosoda} uses soft data augmentation and a momentum mechanism to generate positive and negative samples respectively, utilizing intermodal and intra-modal contrastive learning loss to train their model to find the most similar artifact, given a query artifact.  Similarly, Liu et al. \cite{liu2023graphsearchnet} construct a model called GraphSearchNet, and use bidirectional GNNs to construct graphs for both the code and the queries. Zeng et al. \cite{zeng2023degraphcs} uses word embeddings from CodeBERT \cite{feng2020codebert} and CodeT5 \cite{wang2023codet5+} with variable-based flow graphs and use a gated graph neural network to model these graphs. On the other hand, Sun et al. \cite{sun2022code} generates translations for the programming language artifacts using a program compiler and a disassembler to generate an instruction sequence for the artifact. 

The CodeBERT model \cite{feng2020codebert}, also working on the code search task, used a pre-trained model trained with two different objectives. This pre-trained model consisted of separate entities for the natural language and programming language sequences, and a unified entity for both types of sequences. A special token was used as an aggregated sequence representation for the two types of sequences. These tokens were the measure of similarity between the two types of sequences. During data pre-processing, 1000 pairs of sequences where picked, both where the pairs are actually linked, and pairs where they were not. All of these pairs' representations had these special tokens to measure their similarities. The model was then trained with a binary classification loss function, with a softmax layer connected to those special tokens. In other words, the model was pre-trained on different objectives to generate special tokens that represented the similarity of the sequences. The aggregated sequence, including both sequences and the special tokens were passed on to a modified version of the model, where the special token was used to classify the sequences as either linked or not. The same model is reused by Lu et al. \cite{lu2021codexglue} to evaluate a more constrained version of the Python division of the CodeSearchNet dataset. This dataset, called the ``AdvTest" set, consists of training, validation and testing sets based on the Python division, with many of its rows filtered through based on certain constraints. The chief constraint being the removal of rows where the code could not be compiled into Abstract Syntax Trees, and replacing function and variable names with special tokens. Details on this dataset are discussed later.

Guo et al. \cite{guo2021graphcodebert} builds upon these works by also utilizing the data flow from the programming language sequences during pre-training. This pre-trained model, GraphCodeBERT, is then used for the task of code search on the CodeSearchNet dataset. On the other hand, Wang et al. \cite{wang2021syncobert} pre-trains the SynCoBERT with natural language artifacts, programming language artifacts, and AST representations of the programming language artifacts with different pre-training objectives with a goal of encoding the symbolic and syntactic information of programming languages. Similarly, Guo et al. \cite{guo2022unixcoder} pre-trains the UniXcoder model with natural language artifacts, and flattened sequence derived from the AST representations of the programming languages. In other words, this work pre-trains their model using sequential representations of the ASTs of the programming language artifacts, rather than the programming language sequences directly. Parvez et al. \cite{parvez2021retrieval} makes use of the CodeSearchNet dataset for code generation and summarization with their SCODE-R model. An earlier step in that process included code retrieval or code search. This work also made use of two pre-trained encoders, one for the natural language sequences and the other for the programming language sequences. The retriever module in this work adopts Dense Passage Retriever (DPR) models \cite{karpukhin2020dense}, and the encoders in this module are initialized from GraphCodeBERT \cite{guo2021graphcodebert}. Another contemporary work, Salza et al. \cite{salza2022effectiveness} pre-trains separate BERT encoders, one on natural language sequences, and the other on programming language sequences. The representations learned by these encoders are then used with a Multimodal Embedding Model (MEM) to calculate the similarity of the sequences, and update the two pre-trained encoders' learning. The CodeBERT model \cite{feng2020codebert}, as well as the models following it, GraphCodeBERT \cite{guo2021graphcodebert} and SynCoBERT \cite{wang2021syncobert} follow a similar process of pre-training a model with multiple encoders with some objective to encode the sequences and generate a special token for each group of input sequences, and then fine-tuning this model and using the special tokens to classify the given group of input sequences as either linked to each other or not.

In the pre-training stage, two separate language models are used to generate embeddings for the natural language and programming language sequences. These embedded sequences are concatenated and formatted to include certain tokens. The $[SEP]$ token separates the two sequences, the $[EOS]$ denotes the end of the sequence, and the $[CLS]$ token functions as a representation of the two sequences' similarity. These concatenated sequences are passed as input to train the the CodeBERT model with an objective. The output of the CodeBERT model is a sequence that includes encoded representations for the natural language and programming language sequences, as well as an encoded $[CLS]$ token. During training for the task of code search, the language models are discarded, and a softmax layer is attached to the representation of the $[CLS]$ token. The CodeBERT model is then trained with a binary classification loss function to predict whether the given pair is linked or not. Before testing, the dataset is formatted to form pairs of natural language and programming language artifacts. For each truly linked pair, 999 distractor pairs are formed, with different programming language artifacts, none of which are actually linked to the one natural language artifact. For training and validation, the numbers of linked and non-linked pairs are balanced. Either the natural language artifact or the programming language artifact can be replaced to form these pairs for both training and validation. However for testing, only the programming language artifact is replaced to simulate retrieving the correct artifact from a pool of 1000 possible artifacts.

GraphCodeBERT \cite{guo2021graphcodebert} and SynCoBERT \cite{wang2021syncobert} follow similar processes of pre-training and the use of $[CLS]$ tokens, but use different forms of inputs. GraphCodeBERT substitutes the use of programming language artifacts with data flow graphs extracted using those programming language artifacts, while SynCoBERT substitutes them with AST representations of the programming language artifacts. Similarly, Guo et al. \cite{guo2022unixcoder} replaces the programming language artifacts with their flattened AST representations and uses the natural language sequences alongside them to pre-train their model. The CodeT5+ model \cite{wang2023codet5+} also pre-trains their model consisting of multiple encoders on several objectives, and make use of special tokens for classification.  Hu et al. \cite{hu2023revisiting} builds a framework, TOSS, that integrates the methods from different past works, notably GraphCodeBERT \cite{guo2021graphcodebert} and CodeBERT \cite{feng2020codebert}, first retrieving multiple candidates before ranking them. This results in higher performance, at the cost of time.

Gu et al. \cite{gu2021multimodal}, on the other hand, uses natural language sequences, programming language sequences, and ASTs derived from the programming language artifacts for training. The training of this work's model, UNI-LCRS, involves using separate encoders for these three types of inputs. All three of these encoders transform the inputs to vectors with the same shapes. The vectors derived from encoding the programming language artifacts and ASTs are summed to generate the joint vectors. The cosine similarity between this joint vector and the query vectors are used to update the encoders during the training process.

To summarize, there have been numerous works in the past that make use of the dual encoder structure for the code search task. While these works show impressive performance, the need for auxiliary information such as ASTs, either available separately in the data, or extracted from the data as a separate step can be negative for the task. Models that rely on training on this data might show poor performance if this data is not readily available. On the other hand, extracting and integrating this information can raise the complexity of the model, requiring more resources and time for both training and evaluation. Moreover, without a sufficient amount of this extra information, the model can potentially overfit on this data, leading to poorer performance overall. In addition, existing works do not make use of the language's semantic characteristics, whereas our approach does consider that for Python. This results in our approach performing well for Python, but not for other languages.

The model in this work takes inspiration from these structures to build a simpler dual encoder architecture, and trains FastText language models on each of the used dataset variations to generate the embeddings, but does not extract and encode any contextual information separately. Some of the baseline models, the architecture of this work, the data used, their results, comparison against baseline models, and the results' implications are discussed in the following sections.

\section{Methodology} \label{methodology}
\begin{figure}[!tbh]
    \centering
    \begin{subfigure}[t]{0.75\textwidth}
        \centering
        \includegraphics[width=\linewidth]{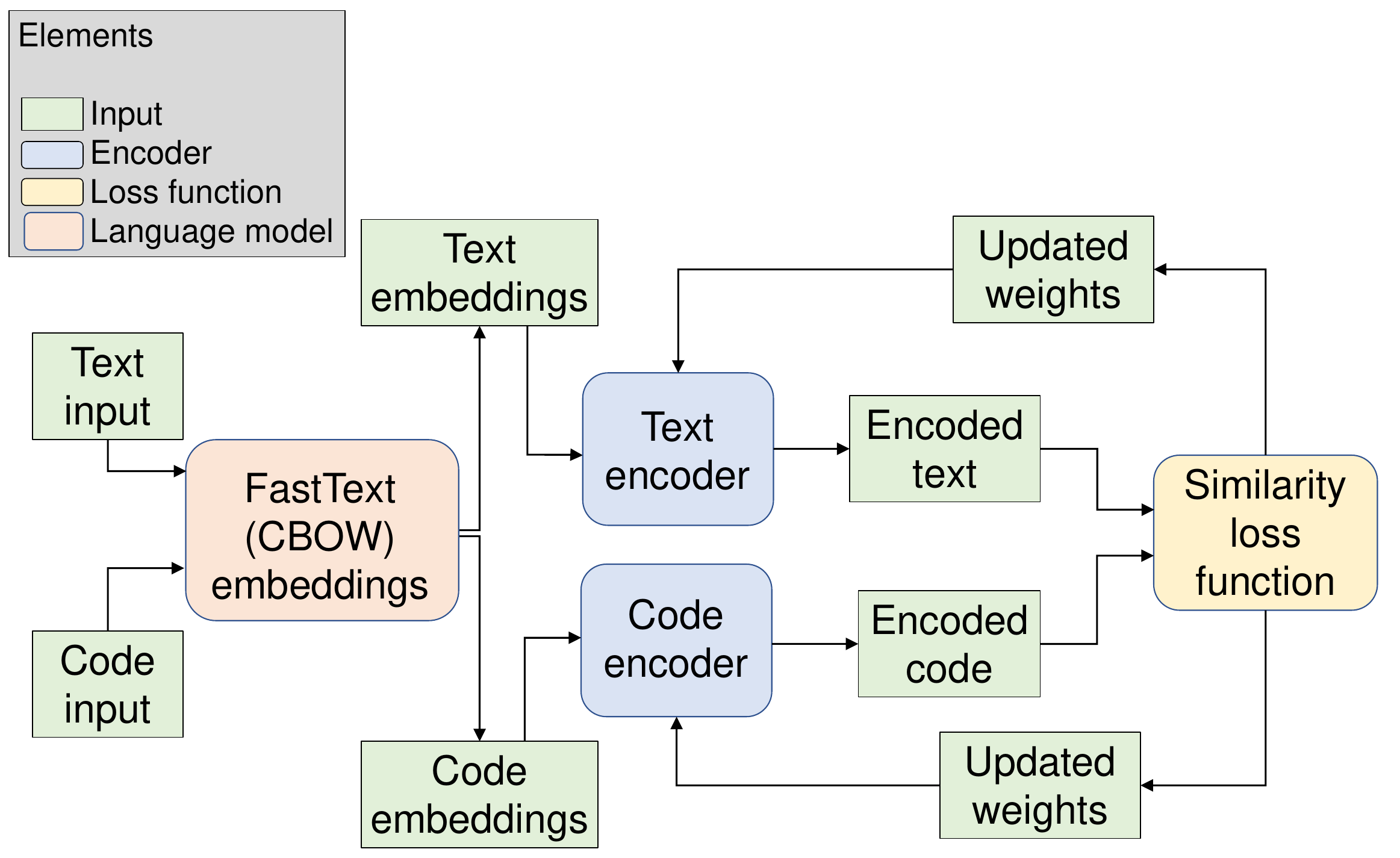}
        \caption{Dual encoder model training pipeline\newline}
        \label{pipeline}
    \end{subfigure}%
    \hfill
    \begin{subfigure}[t]{0.75\textwidth}
        \centering
        \includegraphics[width=\linewidth]{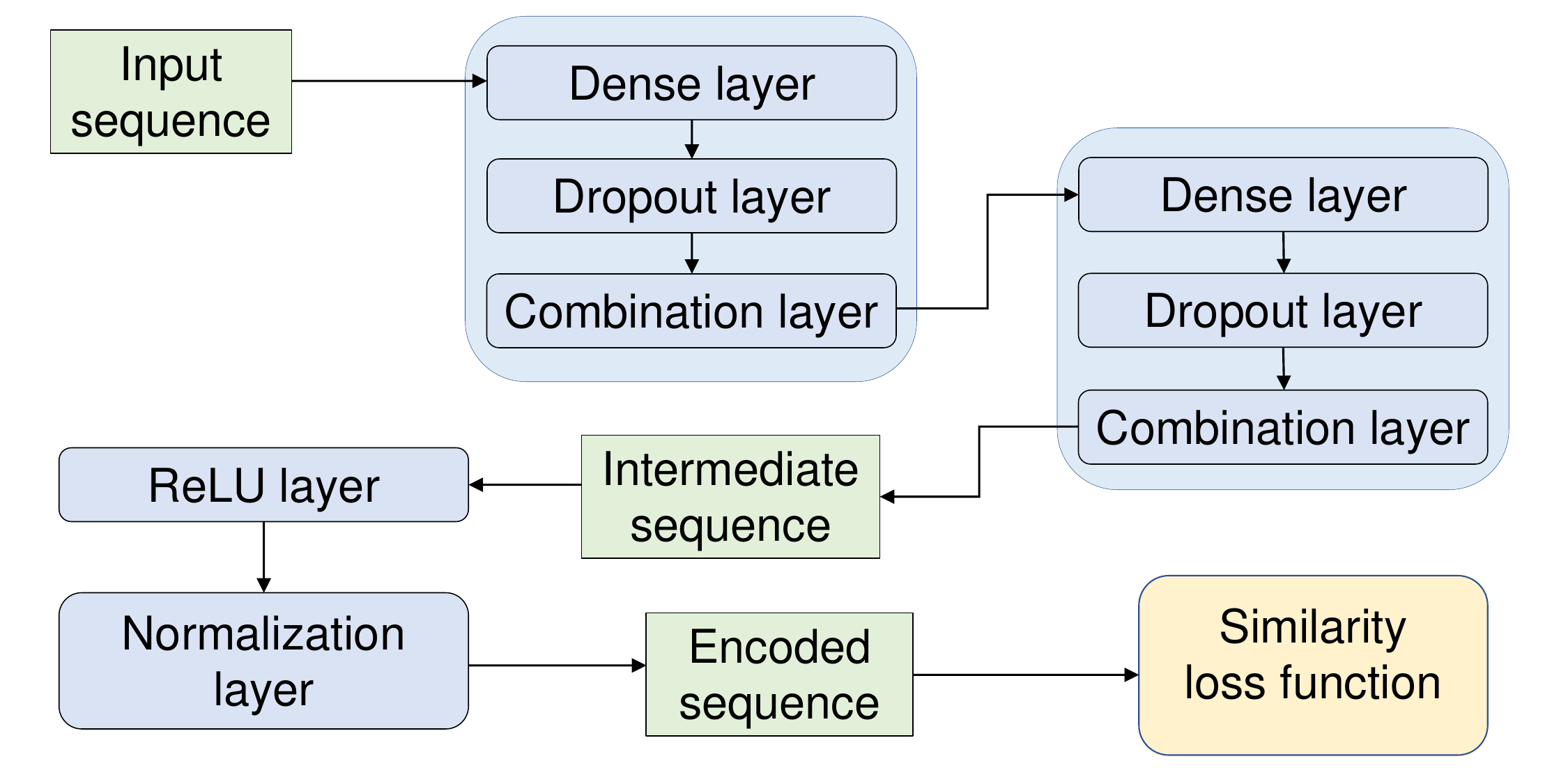}
        \caption{Structure for each encoder}
        \label{encoders}
    \end{subfigure}
    \caption{Pipeline for the proposed approach.}
\end{figure}

Figure-\ref{pipeline} shows an outline of the process here. The raw data undergoes some preprocessing, before being used to train a unified FastText language model. This language model is used to generate embeddings for each input artifact in the data. The embedding pairs are passed on to the dual encoder structure, where the text embeddings are passed on to the text encoder, and the code embeddings are passed on to the code encoder. The structure for each encoder can be seen in Figure-\ref{encoders}. These encoders output encodings, which are used with a cosine similarity loss function to calculate the loss and update the weights for each encoder.

\noindent\textbf{Preprocessing: }All data used here underwent a series of pre-processing steps. These steps included filtering out any pairs containing non-ASCII symbols, removing any non-alphanumeric symbols, and splitting variable or function names into separate words.\\

\noindent\textbf{FastText Embedding: }Text from both the natural language and programming language sequences were then used to train a FastText CBOW (Continuous Bag-of-Words) language model with a 300 dimension size for each of the CodeSearchNet dataset variations. Each of these models were then used to generate the word embeddings for both sequences in each dataset. The data, in this embedded format, was used by the model for training and evaluation.\\

\noindent\textbf{Dual Encoders: }Separate encoders were used for the natural language and programming language sequences. These encoders consisted of three layers with two passes and two following layers as shown in Figure-\ref{encoders}. More specifically, the output from the final layer of an encoder is passed on again to the first layer. The output from the final layer after this second pass is passed on to the final following layers. The repeating three layers consist of a dense layer, a dropout layer, and a combination layer. The final layers are a ReLU (Rectified Linear Unit) layer and a normalization layer. As shown in Figure-\ref{pipeline}, the initial input to these encoders are the word embedding sequences for the natural language and programming language artifacts. After passing through the first dense layer, a sequence then goes through a dropout layer. The dropout layer arbitrarily sets a certain portion of the input sequence's weights to zero. This forces the model to try to extract the necessary information from the remaining portion of the input sequence. Since each pass sets a different combination of points' weights to zero, the final trained model is more efficient at learning from the entire input sequence, rather than specific parts of it. The sequence then moves on to a combination layer. The combination layer combines its input sequence with the current pass's input sequence.

This sequence then moves on to the second pass, which has a similar sequence of layers. After the sequence has passed through this second pass of a dense layer, a dropout layer and a combination layer, it is then passed on to a ReLU layer. The ReLU layer only keeps the non-negative values of its input. This ensures the cosine similarity of the two encoded sequences is always non-negative.

Finally, the input layer is put through an L2 normalization layer which helps simplify the cosine similarity calculation. The output from this layer is the encoded sequence for the given input. An encoded sequence is generated for the natural language and programming language artifacts each. \\

\noindent\textbf{Loss Function: }These sequences are passed onto a cosine similarity loss function to calculate the loss at this stage. In contrast to previous approaches, this loss function does not use 0 as ground truth labels for the non-linked pairs. Given the encoded sequences of a pair of artifacts $(X_i, Y_j)$ the target cosine similarity is 
$$S(X_i, Y_j) = (X_i \cdot X_j + Y_j \cdot Y_i)/2$$
where $X_j$ is a code artifact linked to $Y_j$ and $Y_i$ is a text artifact linked to $X_i$. The binary cross-entropy loss is then calculated:
$$L(X_i, Y_j) = - S(X_i, Y_j)\log (X_i \cdot Y_j) -(1-S(X_i, Y_j))\log(1 - X_i \cdot Y_j).$$
This calculation allows the model to also take into consideration the semantic similarity between each pair into its learning process.\\

\noindent\textbf{Evaluation: }The model is evaluated by feeding it a pair of artifacts. The output from the model is then a similarity score between 0 and 1. For each natural language query, the similarity score with each programming language artifact is first calculated.  These scores are then sorted, and are iterated through. During each iteration, the programming language artifact for the current score is treated as a resultant artifact, and it is checked whether this pair exists in the list of all correct pairs. If it does, then the reciprocal rank (RR) is calculated for that query. The average of these RR scores is the final Mean Reciprocal Rank (MRR) score.

In summary, the factor that separates our work from previous work is not any single change or substitution, but rather the combination of several factors that our approach constitutes of. Nonetheless, key traits of our approach that separate it from previous works can be seen in Table-\ref{difference}.

\begin{table}[!tbh]
    \centering
    \caption{Difference in our approach and past works}
    \begin{tabular}{c|c|c}
    \hline
        Factor & Our approach & Past works\\ \hline
        \makecell{Language\\ model} & Unified FastText model & \makecell{CodeBERT \cite{feng2020codebert}, RoBERTa \cite{feng2020codebert},\\ GraphCodeBERT \cite{guo2021graphcodebert},\\ SynCoBERT \cite{wang2021syncobert},\\ CodeT5+ \cite{wang2023codet5+}, others.}\\ \hline
        \makecell{Sub-word level\\ embedding} & Yes & No\\ \hline
        \makecell{Loss\\ function} & \makecell{Cosine similarity-based loss\\ and Binary cross-entropy loss} & \makecell{Binary classification loss \cite{feng2020codebert}\\ Contrastive loss \cite{shi2023cocosoda}}\\ \hline
        Task framing & Retrieval task & Classification task\\ \hline
        \makecell{Word embedding\\ generation} & \makecell{As a single step before\\ training and testing} & While training and testing\\ \hline
        \makecell{Additional training\\ information (aside\\ from natural and\\ programming language\\ sequences} & No & ASTs \cite{gu2021multimodal} \cite{wang2021syncobert} \cite{guo2022unixcoder}, Data flow \cite{guo2021graphcodebert}\\ \hline
        Testing pool & \makecell{Both entire pool\\ and 999 distractors} & \makecell{999 distractors\\ \cite{gu2021multimodal} \cite{feng2020codebert} \cite{lin2021traceability} \cite{salza2022effectiveness},\\ Both entire pool and\\ 999 distractors \cite{guo2021graphcodebert},\\ Entire pool only\\ \cite{hu2023revisiting} \cite{wang2023codet5+} \cite{shi2023cocosoda} \cite{wang2021syncobert} \cite{parvez2021retrieval}}\\
    \hline
    \end{tabular}
    \label{difference}
\end{table}

 These defining factors are discussed in more detail below -
\bi
    \item The training of a unified language model for both the natural language and programming language artifacts. Using a unified language model for both type of sequences allows for similar artifacts to have similar embeddings, more so if there is an overlap of words between artifacts. Our model uses this as a starting point to learn which artifacts are different, rather than which ones are similar.
    \item The use of a cosine similarity based loss function during training. The use of this loss function directly helps the model in learning which artifacts are similar and which ones are dissimilar. This eliminates the need for separately generating non-linked artifact pairs, as that would be an essential step in training a model to predict whether a pair is linked or not when a classification-based loss function is used.
    \item The use of language models to generate word embeddings as a step before training. This step can prevent the need for generating embeddings while training, which might become time and resource consuming for large amounts of data.
\ei

\section{Experiment setup}
\subsection{Data}
\label{data_subsection}

\begin{table}[!tbh]
    \centering
    \caption{Dataset sizes after filtering}
    \begin{tabular}{p{0.29\linewidth}|p{0.15\linewidth}|p{0.15\linewidth}|p{0.15\linewidth}|p{0.08\linewidth}}
    \hline
         Dataset & Training set & Validation set & Testing set & Total\\ \hline
         CodeSearchNet (Python) & 412,178 & 23,107 & 22,176 & 457,461\\
         AdvTest & 250,680 & 9,562 & 19,113 & 279,355\\
         DGMS & 327,576 & 81,894 & 1000 & 410,470\\ \hline
    \end{tabular}
    \label{dataset_size}
\end{table}

The Python division of the CodeSearchNet corpus, and two of its variations were used as the data for training and testing the model here. The size of the splits in each of the datasets can be seen in Table-\ref{dataset_size}. The Python division of the CodeSearchNet corpus is the only Python-based large dataset used in training and evaluation in past relevant works. The original corpus consists of lines of code in the form of fully coded functions in Python, and their corresponding documentation. The CodeSearchNet task divided these documentation-code pairs into training, validation, and testing sets. Those same sets were also used for the model here. The approach was evaluated in two different ways using this dataset - one  with  1000 artifacts while querying, and one where the entire testing set was used for each query. The dataset where the results with 1000 artifacts while querying is labeled as ``CodeSearchNet Python (Limited)", while the dataset with the complete testing set is labeled as ``CodeSearchNet Python (Full)".

The same training-validation-testing splits for the CodeSearchNet dataset is maintained throughout both this work and the baselines this work is compared against. The limited CodeSearchNet data tests on similar data - all of the queries from the testing split. It is limited by the number of distractor artifacts chosen alongside the correct option. These distractors are chosen at random. But these experiments are conducted a number of times to ensure consistency.

AdvTest is a more constrained version of the CodeSearchNet dataset's Python division, introduced in Lu et al. \cite{lu2021codexglue}. This data filters out any pairs where the programming language artifacts cannot be parsed into ASTs, filters out pairs where any artifact in a pair is either empty or not in English, and filters out pairs based on their lengths and the presence of certain phrases or tokens. This dataset also replaces all function and variable names in the data with special tokens, such as replacing function names with ``$func$", and variable names with ``$arg_i$". This replacement is done only on the validation and testing splits of the dataset.

The DGMS data is a version of the CodeSearchNet Python data used in Ling et al. \cite{ling2021deep}. This version of the dataset underwent different processing steps than AdvTest. Rather than using the pre-divided splits, this version combines these splits, then extracts the docstrings from the programming language artifact in every pair. Then these docstrings are used as the natural language artifacts instead of the original natural language artifacts. Afterwards, any pairs with fewer than three words in the docstring or fewer than three lines of code in the programming language artifacts are filtered out. This combined and filtered collection of pairs are then shuffled and split into training, validation and testing sets, where the testing set contains only 1000 pairs. 80\% of the remaining data is used as the training set here, while the remainder of the dataset is used as validation data. All three of the datasets that were used underwent a series of aforementioned pre-processing steps before being used by the model for training.

\subsection{Word embeddings}
\label{word_embeddings}
FastText word embeddings \cite{fasttextwordembeddings} were used in generating the word embeddings for both the natural text and code sequences due to their ability to capture sub-word level information in the embeddings. Retaining sub-word level information is particularly important here since the code could potentially contain words, such as variable or function names that are combinations of other words found in the related natural text artifact. A CBOW (Continuous Bag-Of-Words) version of a FastText language model with a dimension size of 300 was trained from scratch on the text for each dataset. These models were then used to generate the word representation for all pairs in each dataset. Data was passed on to the model in this embedded format.

\subsection{Evaluation metrics}
\label{evaluation_metrics}
To focus on the retrieval aspect of the task, MRR was the primary metric used for evaluation. To calculate the MRR score, for each natural text query, the rank of its corresponding linked code artifact is retrieved. Rank here simply refers to the position of the linked artifact in the sorted list of matching code artifacts. The inverse of the rank is the Reciprocal Rank (RR) for the query. The mean of all the queries is the MRR scores. For $Q$ queries,
\begin{equation}
    MRR = \frac{1}{Q} \sum_{i=1}^{Q} \frac{1}{Rank_i}
\end{equation}

The MRR score here represents how likely it is for the correct artifact to be the top retrieved artifact when querying. This metric was chosen as the only metric when comparing performance since this was the only metric used by all the baseline models. Some other metrics were also used to show the performance of our approach. However, since not all baselines models use these metrics, they were not used when comparing the results. These other metrics include Accuracy, Mean Average Precision(MAP)@1 and Mean Average Accuracy(MAA)@1.

\begin{equation}
    MAP@1 = \frac{1}{Q} \sum_{i=1}^{Q} Precision@1
\end{equation}
\begin{equation}
    MAA@1 = \frac{1}{Q} \sum_{i=1}^{Q} Accuracy@1
\end{equation}
Here, Precision@1 and Accuracy@1 refer to the Precision and Accuracy scores for the retrieved artifact for each given query respectively.

\subsection{Experimental setup}
The training and  evaluation  steps were conducted on a device with an Intel i7-4790 3.6 GHz processor, 32 GB of system memory, and a NVIDIA GeForce RTX 2070 with a memory of 8 GB. The hyperparameter values used while training the model are shown in Table-\ref{hyperparameters}. The models were run for 300 epochs with early stopping on validation loss.

\begin{table}[h]
    \centering
    \setlength\tabcolsep{8pt}
    \caption{Model hyperparameter values}
    \begin{tabular}{c|c}
    \hline
         Hyperparameter & Value\\ \hline
         Initial learning rate & 0.001\\
         Optimizer & Adam optimizer\\
         Loss function & Cosine Similarity loss\\
         Dropout rate & 0.3\\
         Output size & 2000\\
         Maximum number of epochs & 300\\ \hline
    \end{tabular}
    \label{hyperparameters}
\end{table}



\subsection{Comparison}
\label{baselines_subsection}

Performance on the CodeSearchNet Python dataset was compared against several previous works. The results reported in those works are used when comparing our approach against them. These works are discussed at length in a previous section.

Evaluation on the base CodeSearchNet dataset involved testing on both CodeSearchNet Python (limited) and CodeSearchNet Python (Full). Testing on the limited dataset involved on 1000 pairs at a time. During each testing round, 1000 pairs from the testing set were tested, and then removed from the testing pool. After a number of iterations, all of the pairs in the testing set were tested on. The results from these iterations were aggregated as the final results. This helped make the results from these evaluations comparable to other works where only 1000 pairs were tested at a time \cite{feng2020codebert}. This process also ensured the entire testing set was tested on, making the evaluations more thorough. Including these two types of testing ensured consistency with previous works, since some works tested with 1000 possible artifacts for each query, some tested with the entire testing set, and some evaluated their approach in both ways.

In contrast, evaluation on the AdvTest dataset covered the entire testing pool. Expanding the testing set size to its entire pool made the task much more difficult and time-intensive. Our approach's performance on this dataset is also compared against several models. Finally, performance on the DGMS dataset is compared against only Ling et al. \cite{ling2021deep}. The DGMS dataset is limited to 1000 pairs in its testing set. So the entire dataset was shuffled, split, and used by the model a number of times to generate and report the final mean results.

\section{Results}
\label{results_section}
\begin{table}[h]
    \centering
    
    \caption{Results}
    \begin{tabular}{c|c|c|c|c}
    \hline
         Dataset & Accuracy & \makecell{MAP@1} & \makecell{MAA@1} & MRR\\ \hline
         \makecell{CodeSearchNet Python (Limited)} & 0.946 & 0.888 & 0.919 & 0.919\\
         \makecell{CodeSearchNet Python (Full)} & 0.692 & 0.692 & 0.692 & 0.762\\
         AdvTest & 0.529 & 0.518 & 0.518 & 0.597\\
         DGMS & 0.914 & 0.914 & 0.914 & 0.939\\
    \hline
    \end{tabular}
    \label{results}
\end{table}
Table-\ref{results} shows the performance of our model on the different dataset variations using the evaluation metrics outlined in Section-\ref{evaluation_metrics}. The performance of our approach is evaluated primarily using MRR scores. Numerous works have evaluated their models on the CodeSearchNet dataset in the past. Our approach's performance, in comparison to state-of-the-art models and other baselines models can be seen in Table-\ref{comparison}. The MRR scores represent how likely it is for the actual result to be retrieved as the most similar artifact for a given query. For example, a MRR score of 0.9186 can be interpreted as the model retrieving the correct artifact as the most similar code artifact for a given text query 91.86\% of the time.

\begin{table}[h]
    \centering
    \caption{Comparative results using MRR scores}
    \renewcommand{\arraystretch}{1.2}
    \begin{tabular}{p{0.25\linewidth}|p{0.15\linewidth}|p{0.15\linewidth}|p{0.15\linewidth}|p{0.08\linewidth}}
    \hline
         \multirow{2}{*}{Model} & \multicolumn{4}{c}{Dataset}\\ \cline{2-5}
         & CodeSearchNet (Limited) & CodeSearchNet (Full) & AdvTest & DGMS\\ \hline
         \textbf{Our approach} & \textbf{0.9186} & \textbf{0.7616} & \textbf{0.5967} & \textbf{0.9385}\\
         DGMS\cite{ling2021deep} & - & - & - & 0.922\\ 
         \vspace{-2ex}  TOSS\cite{hu2023revisiting} & \vspace{-2ex}  -  & \vspace{-2ex}  0.759 & \vspace{-2ex}  -  & \vspace{-2ex}  -  \\
         CodeT5+\cite{wang2023codet5+} & - & 0.758 & 0.447 & -\\
         CoCoSoDa\cite{shi2023cocosoda} & - & 0.757 & - & - \\
         \vspace{-2ex}  GraphSearchNet \cite{liu2023graphsearchnet}  & \vspace{-2ex}  -  & \vspace{-2ex}  0.739  & \vspace{-2ex}  -  & \vspace{-2ex}  -  \\
         SynCoBERT\cite{wang2021syncobert} & - & 0.724 & 0.381 & -\\
         GraphCodeBERT\cite{guo2021graphcodebert} & 0.879 & 0.692 & - & -\\
         Uni-LCRS\cite{gu2021multimodal} & 0.8707 & - & - & -\\
         SCODE-R\cite{parvez2021retrieval} & - & 0.690 & - & -\\
         CodeBERT\cite{feng2020codebert} & 0.8685 & 0.672 & 0.507 & -\\
         T-BERT\cite{lin2021traceability} & 0.851 & - & - & -\\
         RoBERTa\cite{feng2020codebert} & 0.8087 & 0.610 & 0.419 & -\\
         CodeSearchNet challenge\cite{husain2019codesearchnet} & 0.6922 & - & - & -\\
         Salza et al.\cite{salza2022effectiveness} & 0.3069 & - & - & -\\ \hline
    \end{tabular}
    \label{comparison}
\end{table}

\begin{figure}[h!]
    \centering
    \includegraphics[width=0.9\linewidth]{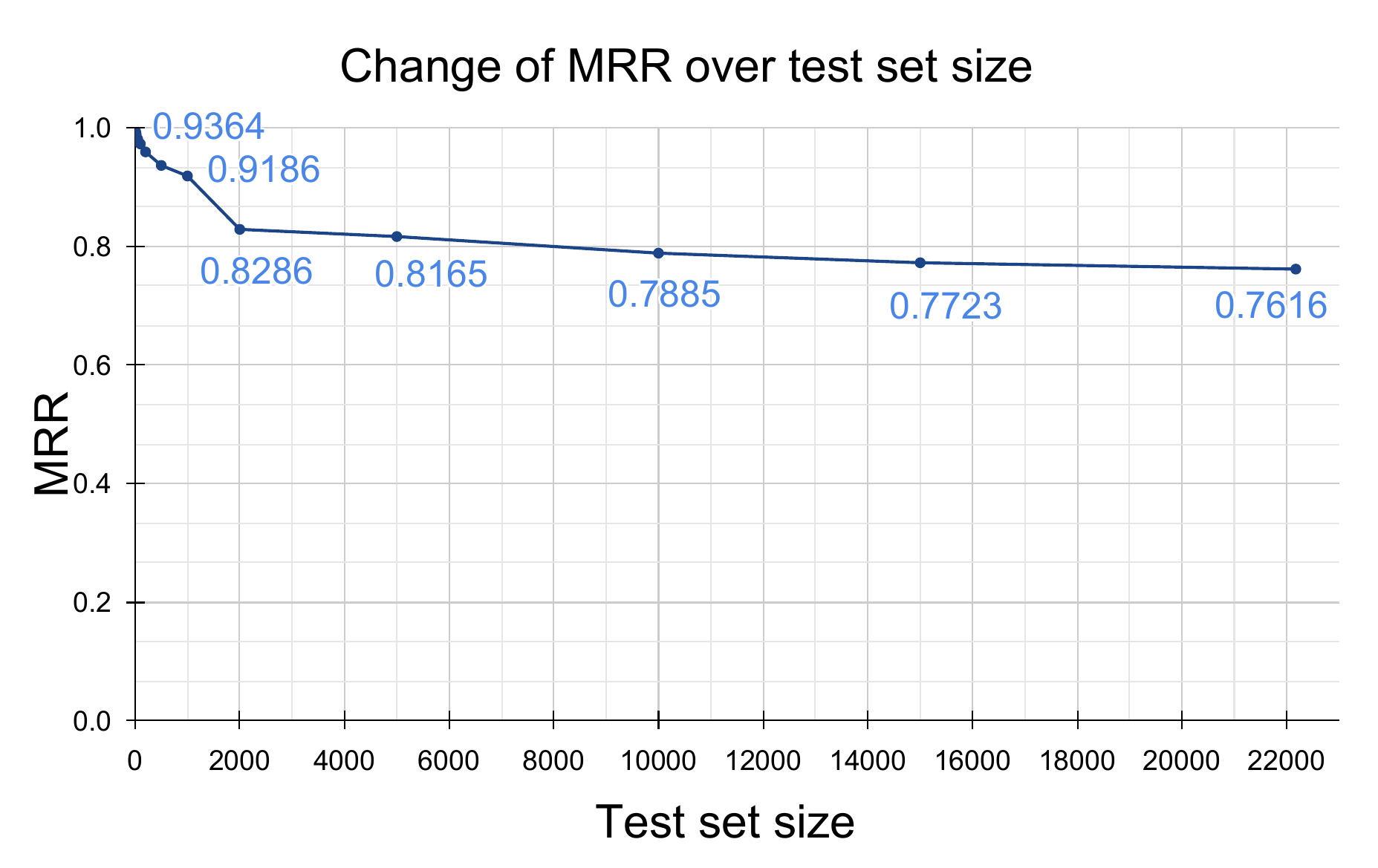}
    \caption{Test set size size vs MRR score for CodeSearchNet Python}
    \label{testsetmrr}
\end{figure}

For the CodeSearchNet dataset, the model was also evaluated with different test sizes between 1000 and the complete testing set.  A change of MRR scores over test set size can be seen in Figure-\ref{testsetmrr}. The fewer distractor artifacts the model has to chose from during testing, the lower the probability of picking the wrong artifact. Moreover, the rank of a correct artifact would be higher even in the event of an incorrect retrieval. This in turn would lower the RR score for each query. With a very small pool of possible artifacts, the model performs almost perfectly. The MRR scores see a significant drop once the test set size crosses 1000. These scores keep declining until the test set size reaches the size of the entire test set of the CodeSearchNet (Limited) dataset.

\subsection{RQ1: Effectiveness}
\label{rq1}
Table-\ref{comparison} showcases our approach's, and other relevant models' results on the code search task in terms of MRR scores. Several of these past works make use of either pre-trained models where the two types of artifacts are used to train separate language models, such as RoBERTa \cite{feng2020codebert} and CodeBERT \cite{feng2020codebert}. In contrast, our approach trains a unified language model incorporating both the natural language and programming language artifacts. While this raises the similarity for non-linked pairs initially, the model is then able to learn which parts of each sequence is important for linked pairs with the help of a similarity loss function.

On the other hand, numerous other works extract contextual information from sequence pairs and include them during training. For example, GraphCodeBERT \cite{guo2021graphcodebert} makes use of the data flow from the programming language artifacts, and Uni-LCRS \cite{gu2021multimodal} encodes the AST representations of the programming language sequences in addition to the sequence pairs themselves. Furthermore, several past works generate pairs where the artifacts are not linked, and make use of a classification loss function to predict whether a given pair is linked \cite{feng2020codebert}.

Both incorporating contextual information separately, and generating these negative pairs separately for training can be intensely resource-intensive. CodeBERT requires 10 hours and 2 hours each for pre-training on each objective, while their best model combines these two objectives. GraphCodeBERT reports requiring 83 hours to pre-train the model. In comparison, our approach takes about an hour to train and generate the word embeddings, and 2 hours to train on CodeSearchNet Python, yet shows higher MRR scores.


\begin{RQ}
\textbf{Answer to RQ1: }Our approach shows performance better than state-of-the-art models for the CodeSearchNet Python, AdvTest and DGMS datasets. Furthermore, through the use of a unified language model, and a similarity loss function which eliminates the need for generating negative pairs--- our approach proves to have a less complex structure, and is less resource intensive. This approach also completes its training faster than state-of-the-art models. A combination of all of this points to the answer for the first research question--- the architecture of our approach is indeed more effective at code search.
\end{RQ}


\subsection{RQ2: Data and Model's roles in effictiveness}
\label{rq2}
Our approach shows much better performance with CodeSearchNet Python when compared to the other datasets. This occurrence can be explained by looking at the dataset's characteristics. CodeSearchNet Python contains a significant amount of pairs where the natural language sequence is partially or completely present within the programming language sequence in the form of the docstrings or comments. Having more words in common between artifacts in linked pairs can help boost the similarity scores of those pairs.  In other words, the model finds it easier to identify linked pairs based on the words or parts of words the artifacts have or do not have in common.  A look at the most frequent words in Table-\ref{overlap} shows that there is indeed a significant overlap among the natural language and programming language artifacts' most frequent words  for CodeSearchNet Python.

However, there might also be a high similarity between non-linked pairs as long as they have words or phrases in common between their artifacts, as it can be seen in Table-\ref{overlap_example}. Aside from variable or function names, it might be likely that multiple pairs would have an overlap of words in their natural language artifacts. And if the programming language artifacts for those pairs contain their natural language counterparts, this overlap might carry on there as well. The model's objective would then be to extract the relevant information that makes linked pairs linked, and not only rely on an overlap of words. Table-\ref{avg_sim} shows that the model does exactly this. Before training, both linked and non-linked pairs have high average similarity scores, although the linked pairs' scores are higher. However, after training, the average similarity score for non-linked pairs decrease by a considerable margin while the linked pairs' score remains close to its pre-training value. Therefore, the high similarity scores between linked artifact pairs would not be sufficient for code search without training, as the similarly high scores between non-linked pairs would result in incorrect artifacts being retrieved more often. 

\begin{table}[h]
    \centering
    \caption{Most frequent words  in CodeSearchNet Python}
    \begin{tabular}{c|c|c|c|c|c|c|c|c|c|c}
    \hline
        \textbf{Text} & \underline{the} & \underline{to} & \underline{of} & \underline{a} & \underline{is} & and & param & \underline{for} & \underline{in} & be\\ \hline
        \textbf{Code} & self & \underline{the} & if & \underline{to} & \underline{in} & \underline{is} & \underline{a} & \underline{of} & return & \underline{for}\\ \hline
    \end{tabular}
    \label{overlap}
\end{table}

On the other hand, the AdvTest dataset replaces function and variable names with special tokens. Therefore, the model has to focus on other keywords to capture the similarity between artifacts, without using function and variable names. This is undoubtedly a more difficult task, which is reflected in the evaluation of models on this dataset. The MRR score drops from 0.7616 on the CodeSearchNet Python dataset (Full) to 0.5967 on the AdvTest dataset. In other words, without an overlap of unique words, the model's MRR scores see a drop. However, this model still manages to show better performance than baseline models.  In other words, overlapping words between linked artifact pairs is not necessary for the model's learning, but it is beneficial.

\begin{table}[h]
    \centering
    \caption{Average similarity scores on CodeSearchNet Python (Limited)}
    \begin{tabular}{c|c|c|c}
    \hline
         Trained & Linked pair & Non-linked pair & MRR score\\ \hline
         No  & 0.79 & 0.57 & 0.42\\
         Yes & 0.72 & 0.09 & 0.92\\ \hline
    \end{tabular}
    \label{avg_sim}
\end{table}

In light of all of this, our approach works not by learning why linked pairs are linked, but by learning why non-linked pairs are not linked. In the process, it keeps linked pairs' similarity scores higher, and lowers non-linked pairs' similarity scores significantly. The magnitude of linked pairs' high scores and non-linked pairs' low scores are directly tied into their retrieval scores.  Therefore, the key to improving this model's performance is to increase the linked pairs' similarity scores while decreasing non-linked pairs' similarity scores.  This simpler architecture is appropriate for the task since the model would have two different goals for two different types of input sequences, where each encoder can take on each of those inputs and goals. This helps support the answer to the first research question, and expands on it by explaining in what way the architecture of our approach is efficient at code search.

\begin{RQ}
\textbf{Answer to RQ2: }The results explain why the code search task is easier for models when there is  a high overlap  of words or phrases between linked pairs. Nonetheless, results also show our model being able to learn to differentiate linked and non-linked pairs even when function and variable names are transformed. This proves that our model is capable of extracting contextual information from the pairs to learn their similarities and differences, even if any such information is not provided separately. This conclusion helps answer the second research question of what roles the data and the models play in making the code search task more efficient.
\end{RQ}

\section{Discussion}
\label{discussion_section}

This section outlines some additional experiments conducted, which do not directly help answer our research questions. However, the results from these experiments help support our primary findings, and inform the direction of our future work.

\subsection{Other CodeSearchNet languages}
\label{other_lang_discussion}
\begin{table}[h!]
    \centering
    \caption{Results on other code search languages}
    \begin{tabular}{c|c|c|c|c}
    \hline
         Language & Accuracy & \makecell{MAP@1} & \makecell{MAA@1} & MRR\\ \hline
         Java & 0.382 & 0.381 & 0.381 & 0.506\\
         JavaScript & 0.402 & 0.396 & 0.396 & 0.523\\
         Ruby & 0.387 & 0.369 & 0.369 & 0.507\\
         GO & 0.413 & 0.409 & 0.409 & 0.532\\
         PHP & 0.452 & 0.450 & 0.450 & 0.569\\
    \hline
    \end{tabular}
    \label{results_other_langs}
\end{table}

\begin{table}[h]
    \centering
    \caption{Comparative results on other languages using MRR scores}
    \renewcommand{\arraystretch}{1.2}
    \begin{tabular}{c|c|c|c|c|c}
    \hline
         \multirow{2}{*}{Model} & \multicolumn{5}{c}{Language}\\ \cline{2-6}
         & Java & JavaScript & Ruby & GO & PHP\\ \hline
         \textbf{Our approach} & 0.506 & 0.523 & 0.507 & 0.532 & 0.569\\
         GraphCodeBERT \cite{guo2021graphcodebert} & 0.757 & 0.711 & 0.732 & 0.841 & 0.725\\
         CodeBERT \cite{feng2020codebert} & 0.748 & 0.706 & 0.693 & 0.840 & 0.706\\
         RoBERTa \cite{feng2020codebert} & 0.666 & 0.606 & 0.625 & 0.820 & 0.658\\
         \makecell{CodeSearchNet\\ challenge \cite{husain2019codesearchnet}} & 0.587 & 0.451 & 0.365 & 0.681 & 0.601\\
         Salza et al. \cite{salza2022effectiveness} & 0.291 & 0.311 & - & - & -\\
         Uni-LCRS \cite{gu2021multimodal} & - & - & 0.364 & - & -\\
         \hline
    \end{tabular}
    \label{comparison_others}
\end{table}

Table-\ref{results_other_langs} showcases the proposed model's performance on the other languages within the CodeSearchNet dataset, while Table-\ref{comparison_others} compares these results against baseline works. These experiments were conducted similarly to the CodeSearchNet Python (Limited) experiments, with 1000 pairs tested each time until the entire testing set was exhausted. All of the metrics show much lower scores for the other languages when compared to Python. Some likely reasons behind this difference are discussed below:

\begin{itemize}    
    \item \textbf{Semantic characteristics:} The semantic differences between the six languages is likely the primary reason behind the difference in results. Compared to the other languages, Python-specific keywords are more commonly seen in natural language text than keywords from most of the other languages, as seen in Table-\ref{frequent_words}. In Table-\ref{frequent_words}, programming language words that are also in the top 10 frequent natural language artifact words' list for those languages are highlighted. On the other hand, languages such as Java are much more verbose than Python, with variable and function names often being combinations of different words that are broken into their original words during pre-processing. Both of these characteristics lead to an increased vocabulary size for the non-Python languages, where there are more words with low occurrences. This makes training the word embeddings more difficult compared to Python, where the vocabulary is comparatively smaller with fewer words that have low occurrences. The relatively better word embeddings training for Python is likely a primary reason behind the dual encoders' learning process being smoother and more accurate as well.

    \begin{table}[h!]
    \centering
    \caption{Top 10 frequent words in programming language artifacts}
    \begin{tabular}{|c|c|c|c|c|c|}
    \hline
        Python & Java & JavaScript & Ruby & GO & PHP\\ \hline
        self &  if &  if &  end &  return &  this\\ \hline
        \textbf{\textit{the}} &  return &  this &  if &  err &  if\\ \hline
        if &  new &  0 &  \textbf{\textit{name}} &  if &  return\\ \hline
        in &  String &  function &  def &  \textbf{\textit{nil}} &  function\\ \hline
        \textbf{\textit{to}} &  null &  var &  \textbf{\textit{to}} &  func &  array\\ \hline
        return &  public &  return &  nil &  \textbf{\textit{s}} &  public\\ \hline
        for & 0 &  \textbf{\textit{i}} &  \textbf{\textit{options}} &  string &  new\\ \hline
        is &  int &  the &  new &  \textbf{\textit{v}} &  null\\ \hline
        \textbf{\textit{a}} &  final &  1 &  do &  error &  0\\ \hline
        0 &  this &  \textbf{\textit{options}} &  \textbf{\textit{id}} &  \textbf{\textit{c}} &  value\\ \hline
    \end{tabular}
    \label{frequent_words}
    \end{table}
    \vspace{-1ex}
    Due to Python having more words in its programming language artifacts that are commonly seen in natural language text, and these words being higher up on the list - indicating that these are more frequent in Python than the overlapping words are for Java, JavaScript and PHP, we can deduce this semantic difference to be the primary reason Python performs better than those languages for code search. Ruby and GO have more of these overlapping words than Python. However for GO, most of these words seem to be variable names - which are not very informative for the model to learn from, since they might have different meanings from artifact to artifact. In other words, this high overlap of words between natural and programming language artifacts actually hurts the model's learning process more than it helps. Thus, the semantic difference can be used to also explain the model's worse performance on GO.
    
    \item \textbf{Dataset size:} The CodeSearchNet Ruby and JavaScript datasets are much smaller compared to the rest, as it can be seen in Table-\ref{dataset_size_other}. Since the model is built primarily focusing on Python-based data where the datasets are much larger, that can explain why the model learns and performs poorly when there is less data to learn from. Despite having more overlap in natural and programming language artifacts in Ruby, similar to Python, Ruby has about 11.83\% of the number of training pairs Python does. This much smaller dataset size might be the contributing factor behind the model's poor performance on Ruby.
    \begin{table}[h]
        \centering
        \caption{CodeSearchNet corpus sizes after filtering }
        \begin{tabular}{c|c|c|c|c}
        \hline
             Language & \makecell{Training\\ set size} & \makecell{Validation\\ set size} & \makecell{Testing\\ set size} & Total\\ \hline
             Python & 412,178 & 23,107 & 22,176 & 457,461\\
             Java & 424,451 & 15,328 & 26,909 & 466,688\\
             JavaScript & 123,889 & 8,253 & 6,483 & 138,625\\
             Ruby & 48,791 & 2,209 & 2,279 & 53,279\\
             Go & 317,832 & 14,242 & 14,291 & 34,6365\\
             PHP & 523,712 & 26,015 & 28,391 & 578,118\\ \hline
        \end{tabular}
        \label{dataset_size_other}
    \end{table}
    
\end{itemize}

These differences between the languages call for an approach that is suited to all of these languages. The proposed approach focused only on Python-based data, tweaking and modifying the structure and process based on the nature of the Python-based datasets. Therefore, this approach might not be the most appropriate approach for the other languages. This could explain the relatively poor performance on the other languages when compared to past works. Exhaustive experimentation is included in the outlined future work. This future work would include modification of the structure and pre-processing steps to ensure reasonable performance across all languages.


\subsection{Control experiments}
\label{control_experiments}
\begin{table*}[h]
    \centering
    \caption{Summary of control experiments' results}
    \label{tab:config}
    \begin{tabular}{c|c|c|c|c}
    \hline
         Language model & Loss function & Output size & Number of passes & MRR\\ \hline
         \textbf{Unified} & \textbf{Cosine similarity loss} & \textbf{2000} & \textbf{2} & \textbf{0.9186}\\
         Separate & Cosine similarity loss & 2000 & 2 & 0.01\\
         Unified & Softmax loss & 2000 & 2 & 0.5720\\
         Unified & Contrastive loss & 2000 & 2 & 0.2078\\
          \makecell{Unified (no sub-word\\ level embeddings)}  &  Cosine similarity loss  &  2000  &  2  &  0.8249  \\
         Unified & Cosine similarity loss & 2000 & 1 & 0.8312\\
         Unified & Cosine similarity loss & 2000 & 3 & 0.8259\\
         Unified & Cosine similarity loss & 500 & 2 & 0.8818\\
         Unified & Cosine similarity loss & 8000 & 2 & 0.9107\\ \hline
    \end{tabular}
\end{table*}

The approach was also evaluated through different control experiments where different components of the model were substituted, or certain hyperparameter values were changed. The performance on the model was also recorded for each of these configurations. All these evaluations were done on CodeSearchNet Python (Limited). These evaluations and their results helped in determining the final configuration of our approach and the model. Table-\ref{tab:config} shows a summary of these evaluations, while the following subsections discuss these factors and their effects in detail.

\subsubsection{Language model}
\label{language_model}
As previously discussed, FastText word embeddings were used in this work because of their ability to capture sub-word level information in their generated embeddings. For this work, a unified language model was used, both for the text and the code. The performance was also noted with separate language models for text and code  , and after disabling sub-word level embeddings  . Indeed, this is an approach taken by some previous works \cite{feng2020codebert}. However, the model in our approach performs very poorly when using separate FastText language models. The MRR score while testing on the CodeSearchNet Python (Limited) dataset with a unified language model is 0.9186. This score sees a massive drop to 0.01 when using separate language models.  Disabling sub-word level embeddings also cause a drop in performance, reducing the MRR to 0.8249.

When training a language model, involving both the text and code allows similar embeddings for pairs with a higher overlap of words. This in turn helps the model in understanding the similarity between these artifacts, based on the embedding values. However, having separate language models can result in very different embedding values for similar artifact pairs. The model then has to try to draw connections between these artifacts with contrasting values. With such a large dataset, this task becomes much harder. This leads to poorer learning for the model, and consequently, poorer performance. In other words, with separate language models, the model faces more difficulty in trying to learn about different artifact pairs' similarities and dissimilarities. This is reflected in the similarity score between pairs when testing. While with a unified language model, the similarity scores for linked pairs are much higher than the ones for non-linked pairs, these scores are almost equal when using separate language models.

\subsubsection{Loss function}
\label{loss_subsection}
A cosine similarity loss function was used in the model in this work. The steps for calculating this loss during training are discussed in Section-\ref{methodology}.  A softmax loss and contrastive loss were also included in evaluations. Both of these cases yielded poor results. The nature of the data and its embeddings are the likely reasons behind these poor results. The embeddings for each artifact contain a number of numerical values. These values express no explicit meaning on their own. Rather, the collection of these values is used to represent the artifact. And since the task involves finding similar artifacts, there needs to be some comparison between these artifacts to determine their level of similarity. Although softmax loss might be appropriate for certain classification tasks, the relationship between artifacts does not play a significant role in the calculation of this loss. Softmax loss is therefore, not the most appropriate loss function for this task.

The contrastive loss function shows poor performance for a similar reason. Contrastive loss involves a step of calculating the Euclidean distance between artifact pairs. This distance, and whether the pair is truly linked or not, are used to calculate the loss. Two artifacts might have values with similar proportions but values that are far different from each other. Contrastive loss would show a high loss for this pair, even though they are similar and linked. Moreover, trying to force linked pairs' encodings to have the same values, in contrast to having similarly proportional values is a much more difficult task for the model. Therefore, contrastive loss was not used in the final configuration.

\subsubsection{Output size}
One factor that was noted to influence the artifacts' embeddings was the output size of the model. The output size is the size of the encoded embeddings. The higher this output size, the more information about an artifact the model is able to encode within this encoding. When more information is present, linked pairs will show a higher overlap of this information. This, in turn can lead to better understanding and better performance for the model. However, there is only certain amount of useful information that the model can extract. After certain output size value is reached, these encoded embeddings can contain unhelpful or repeating information which can have a negative impact on the model's understanding.
\begin{figure}[!tbh]
    \centering
    \includegraphics[width=0.75\linewidth]{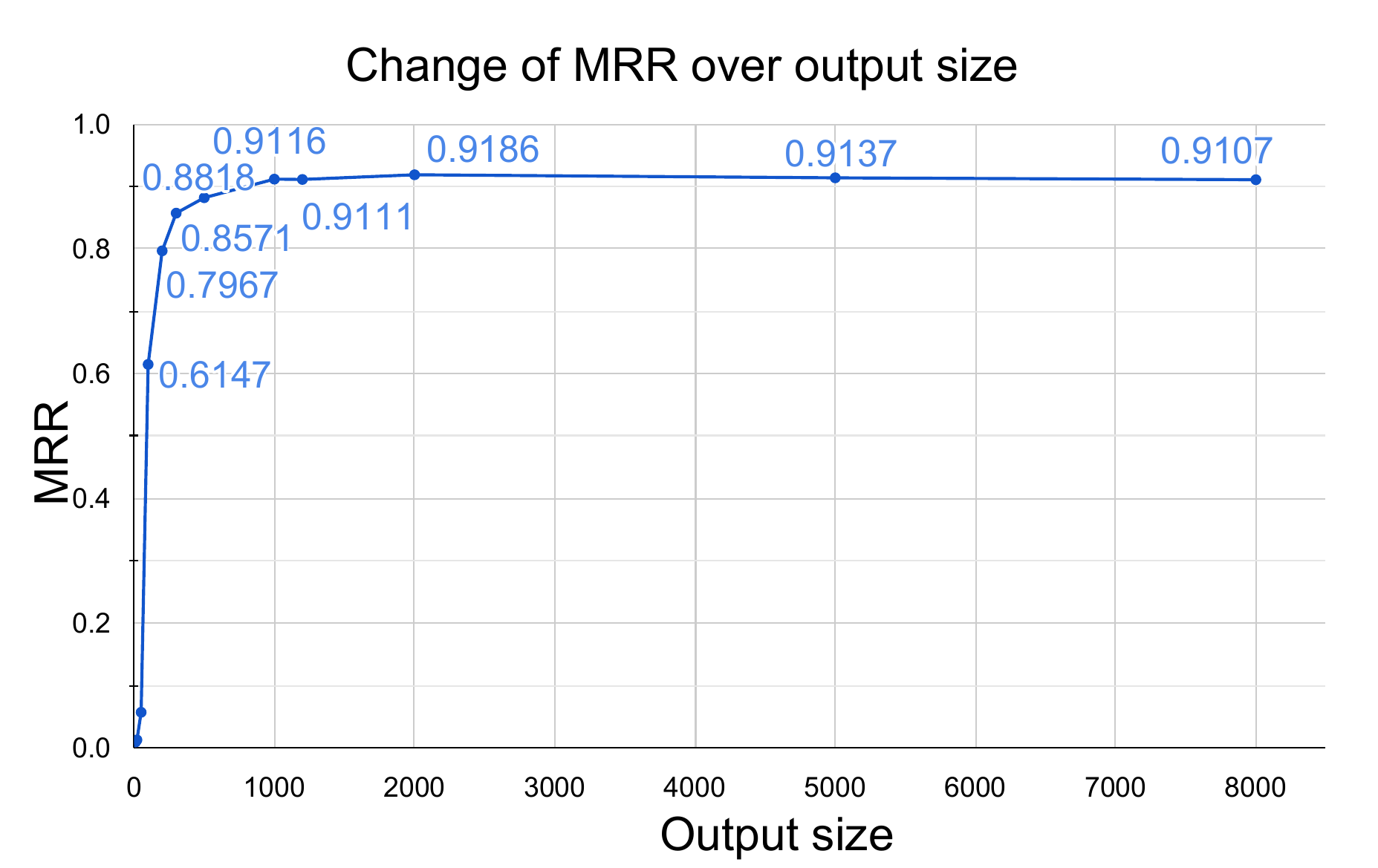}
    \caption{Output size vs MRR score for CodeSearchNet Python (Limited)}
    \label{outputmrr}
\end{figure}

 Figure-\ref{outputmrr} illustrates this relationship in our evaluations . Initially, the MRR scores increase with the output size. But after crossing the 2000 mark, the MRR score starts to decrease. After this point, the model is likely approximating or estimating information to fill out the output size in the encodings. This excess information can lower the similarity between linked pairs' artifacts. Consequently, the model achieves poorer learning, and shows poorer performance. Therefore, an output size of 2000 was picked to balance having the maximum amount of usable information for the model without forcing it to include approximated information.

\subsubsection{Number of passes}
The number of passes for each encoder was another factor that affected the model's learning. With only one pass, the model is not able to fully learn what makes each artifact pair similar or dissimilar. This is reflected through higher training and validation losses during training. On the other hand, with more than two passes, the model overfits on the given data, showing similarly high loss. Having two passes for each encoder ensured the model learns and performs sufficiently well.

\section{Threats to validity}
\label{threats}
Despite having better performance than state-of-the-art models, this work suffers from some drawbacks. Some limitations observed from the data and model's performances are discussed below as different categories of threats to this work's validity -

\bi
    \item \textbf{Internal validity:} A high output size of encoded data is needed for the model to show the reported performance. For very large datasets, generating and storing this large amount of high dimensional data could prove resource-intensive.
    A potential solution to this issue would be thorough investigation into what these high dimensional data represent. A better understanding could lead to a better optimization of this data, or a trade-off between the output size and the performance.
    \item \textbf{External validity:} The proposed model's generalizability with different types of data needs to be explored to ensure the model maintains its performance across different applications. This model's performance was primarily evaluated on Python-based data. Preliminary experiments showed the approach performing poorly for other languages in the CodeSearchNet dataset. Section-\ref{discussion_section} discusses possible reasons behind these performances, as well as potential solutions. Since the vocabulary and structure of other programming languages might be different from Python, the model's learning process might take some different paths as well. This might lead to the model showing different performance for different data. In a similar vein, the proposed model shows lower MRR scores on data where there is less overlap of text between the natural language and programming language artifacts. Even though the performance for these data show higher scores than state-of-the-art models, this performance might not be useful enough in some practical scenarios, such as - if the paired artifacts do not follow consistent naming conventions, or for certain programming languages which are not as intuitive as Python. The model needs to be evaluated using different programming languages to guarantee the model's general high performance.
    \item \textbf{Construct validity:} The proposed model's performance is compared against the reported performances of several other approaches. Although the other approaches use the same benchmark, they can have unmentioned different settings, which could potentially invalidate the results.
    \item \textbf{Reliability validity:} No threats to the work's reliability validity were found.
    \item \textbf{Conclusion validity:} No threats to the work's conclusion validity were found.
\ei

\section{Conclusion and future work}
\label{conclusion}
In conclusion, this work treats the code search task as a multi-language translation task. Through the use of unified FastText word embeddings and cosine similarity based loss on a dual encoder architecture, this work achieves performance better than state-of-the-art models. Through the analysis of the data, the model, and their relationship, this work also highlights certain key aspects of the datasets used in this task. These analyses could help inform future work in the code search task and provide more efficient directions to take while tackling this dataset. This work also serves as a promising sign for the efficient and practical use of dual encoders in other similar tasks.

There is still room for improvement or building upon this work, however. Some future work involving this model could include -

\bi
    \item Deeper investigation into the nature of datasets that the model performs relatively poorly on, as well as the model, and how they affect the performance.
    \item Modifying the training process, or introducing pre-training steps to accommodate for smaller code search datasets, such as the OSS dataset.
    \item Experiments with similar tasks that could be framed as multi-language translation tasks with the use of the dual encoder architecture. These tasks could be in similar contexts with different artifact, even non-text based ones. 
    \item The high performance of the model where both artifacts have a larger overlap of natural text could help with the model's use in tasks from very different contexts. For example, for the task of finding relevant job applications, an ideal application would have a higher overlap of words or phrases with a job posting if some of the applicant's qualifications match with the job posting. In other words, if the applicant meets some of the requirements put forth by the job posting, the job application will have some words or phrases in common with that posting. Since the dual encoder model has been observed to perform better for cases where there is such a high overlap of words, this task and other similar ones might prove ideal applications of this model.
\ei

\section{Data Availability}
The code and link to the data for our approach here are available at \url{https://github.com/hil-se/CodeSearch}. This repository includes links to both the raw text data without any pre-processing, as well as the pre-processed data as word embeddings.



\begin{acknowledgements}
Funding in direct support of this work: NSF grant 2245796.
\end{acknowledgements}

\newpage

%
%

\bibliographystyle{spmpsci}      
\bibliography{bibliography2}   


\clearpage
\appendix

\end{document}